\documentstyle[12pt]{article}

\begin{document}

\newcommand{\Tr}{\mathop{\rm Tr}}
\def\sfrac#1#2{{\textstyle\frac{#1}{#2}}}  

\begin{titlepage}
\begin{center}

\hfill\today\\
\hfill HUTP-95/A043\\
\hfill hep-ph/9511307\\

\vskip .5in
{\large\bf
The Derivation of $M_{\rm GUT} =10^{16}$ GeV\\
from A String Model} \\
\vskip .2in
{\bf Rulin Xiu%
\footnote{\rm E-mail: {\tt xiu@huhepl.harvard.edu}}} \\

{\em Lyman Laboratory of Physics \\
    Harvard University\\
      Cambridge, MA 02138}

\end{center}

\vskip .5in

\begin{abstract}
We propose that gaugino condensation in the hidden sector may
dynamically induce intermediate gauge symmetry breaking and specify
the compactification scale.
We also show that this scheme makes possible affine level one
grand unified string models with gauge breaking scale
$M_{\rm GUT} \sim 10^{16}$ GeV.
$T$-duality plays a crucial role in making this scheme possible,
even though it is spontaneously broken.
We also discuss the generation of the large mass hierarchy between the
string scale and the electroweak scale  and the solution
of the dilaton runaway problem in this scheme.

\end{abstract}
\end{titlepage}

Precision electroweak measurements indicate that supersymmetric
grand unified models lead to a good agreement with a single unification
scale of $M_{\rm GUT} = 10^{16 \pm 0.3}$~GeV~\cite{gut1} and a best fit for
supersymmetry breaking scale
$M_{\rm SUSY} \simeq 1$~TeV~\cite{gut1,gut2,gut3}.
In this paper, we explore the possibility of deriving
these features from string models.

In string theory, all gauge interactions and gravity are unified
\cite{string1,string2}.
This unification happens at around the string scale, which is related to
the gauge coupling constant at the string unification scale through
the relation: $M_{\rm S}^2 = \frac{1}{2} g^2 M_{\rm P}$ \cite{kap,tom}.
Here $M_{\rm P} = 1 / \sqrt{8 \pi G_{\rm N}}=10^{18}$~GeV is the
reduced Planck mass, where $G_{\rm N}$ is Newton's constant.
This indicates that the string gauge coupling constants unify
at $10^{17}~{\rm GeV}$--$10^{18}$~GeV for affine level one string models
without string threshold corrections.
It is not possible to obtain a scale of $10^{16}$~GeV without
invoking string threshold correction and
fine-tuning the gauge coupling constant and the string threshold
correction \cite{con}.

Given this situation, it is natural to consider string models with
grand unified groups which break below the string scale.
However, it is not known how to dynamically induce intermediate scale
gauge symmetry breaking in string models.
In fact, it has been shown that adjoint Higgs representations of low-energy
gauge groups cannot exist in affine level one string models, the simplest
and most well-studied class of models.
This means that the usual breaking mechanisms from the grand unified group
$SU(5)$ or $SO(10)$
to the standard model gauge group does not work in these models.
Adjoint Higgs representations can exist in more complicated string
models, for example higher affine level string models.
However, it is much more difficult to build ``realistic'' models of
this type.
In the last few years, much effort has been devoted to this problem,
with no convincing sucess so far.

In this letter, we propose that intermediate gauge symmetry
breaking at $10^{16}$~GeV may be induced by gaugino condensation
in the hidden sector.
In particular, we propose that some charged
background field VEVs in the observable $E_8$ sector
can be turned on when gaugino condensation happens in the hidden sector,
and that this can lead to the intermediate gauge symmetry breaking.
We find that this stringy gauge symmetry breaking scheme
makes affine level one grand unified string models possible.
The resulting supersymmetry breaking scheme has some advantages
over traditional approaches \cite{wit2} and may even lead to
determining the dilaton at a realistic value
from gaugino condensation dynamics.

In the following, we will review some important features of
gaugino condensation in the heterotic string theory before
describing our scheme.
In the end, we will
discuss about generating the large mass hierarchy between
$M_{GUT}=10^{16}GeV$ and $M_{SUSY}=1TeV$ in this kind
of schemes.

Heterotic string theory has a hidden sector with gauge group $E_8$.
This $E_8$ (or its unbroken subgroup) is asymptotically free and will
become strong at a ``confining'' scale $\Lambda$.
One expected result of this confinement is that the gaugino bilinear
will get a VEV.
{}From the 10-dimensional effective lagrangian of the heterotic string,
one can see that the gaugino bilinear couples linearly to the
antisymmetric tensor field strength, defined as \cite{heter}
\begin{equation}
H = dB - \omega_{3Y} + \omega_{3L},
\end{equation}
where $B$ is the 2-index antisymmetric tensor field, and $\omega_{3Y}$ and
$\omega_{3L}$ are the Yang-Mills and Lorentz Chern-Simons symbols:
\begin{eqnarray}
\omega_{3Y} &=& \frac{1}{30} \Tr(A F - \sfrac{1}{3} A A A),
\\
\omega_{3L} &=& \frac{1}{30} (\omega R - \sfrac{1}{3} \omega \omega \omega).
\end{eqnarray}
It was observed in \cite{wit2} that the terms in the lagrangian that depend on
$H_{ijk}$ and the gauge-invariant gaugino bilinear
$\bar{\lambda} \Gamma_{ijk} \lambda$ combine into a perfect square:
\begin{equation}
\delta{\cal L} \sim (H_{ijk}-
\sqrt{2} g_{10}^2 \phi^{3/4} \bar{\lambda} \Gamma_{ijk} \lambda)^2.
\end{equation}
This means that when the gaugino bilinear obtains a VEV, the dynamics forces
$H_{ijk}$ to obtain a VEV as well.

Various schemes have been proposed in which the VEV of $H$ arises from
different sources.
In ref.~\cite{wit2}, it is proposed that
$\langle H_{ijk}\rangle = c \epsilon_{ijk} =\langle dB \rangle$.
Unfortunately, in this case $c$ is integer quantized in planck units
\cite{rom}, so gaugino condensation is forced to take place near the Planck
scale.
It is  proposed in refs.~\cite{subr1,ross} that the necessary constant term
comes from matter field VEVs.
In this case, the $H$ is not quantized.

In this letter, we propose a modified version of the mechanism proposed in
refs.~\cite{wit2,subr1,ross}.
In particular, we assume that the induced $H$ VEV
arises from the Chern--Simon term rather
than from the  antisymmetric tensor strength or matter fields.
Furthermore, to satisfy the equation of motion, we require that
the induced Chern--Simon term satisfies the condition:
$d\omega=F_{ij}F^{ij}=0$, {\em i.e.}  $F_{ij}=0$.
It appears that in all known string models
\begin{equation}
\label{quant}
\frac{1}{2\pi}\int_{\tilde{W}}H = n + \frac{1}{m},
\end{equation}
where $n$ and $m$ are integers.
However, we will discuss a possible mechanism to generating a small
value for $H$ below.

We now explain how the above proposal may make affine level one
grand unified models possible.
Since the induced charged background VEV's do not correspond to
any four-dimensional physical modes, they can be in the adjoint
representation of the observable gauge interaction $E_6$.
They can break the grand unified string models to low-energy
standard-like models  through the string Higgs effect \cite{hig1}.
Specifically, the VEV's of the charged background fields give masses
to the matter fields $\Phi$ via a cubic superpotential of the form
$W \sim A\Phi\Phi$ and break the gauge symmetry to the gauge subgroup
that commutes with $\langle A \rangle$.
In this case, the masses of the matter fields as well as the gauge
symmetry breaking scale will be of order $\langle A \rangle$.
Since it is determined by gaugino consensation dynamics, the VEV of $A$
will be
\begin{equation}
\langle A \rangle \sim M_{\rm S} e^{-S / 2b_0}.
\end{equation}
We fix the dilaton VEV phenomenologically at $\langle S \rangle \simeq 2$,
the value consistent with the observed weak scale couplings.
For affine level one string models with unbroken $E_6$ hidden sector
gauge symmetry, we have $b_0 = 36 / (16\pi^2)$, which gives
$\langle A \rangle \sim 10^{16}$~GeV.
(For $E_8$ hidden sector, we have $b_0 = 90 / (16\pi^2)$, which gives
$\langle A \rangle \sim 10^{17}$~GeV.)
This scale can be identified with the grand unification scale $M_{\rm GUT}$
from the low-energy point of view.
The heavy massive fields generated in this mechanism will not
substantially alter the weak-scale predictions, because these masses are
all on the order of $M_{\rm GUT}$.
One can also avoid the doublet-triplet problem
 by using the the pseudo-Goldstone schemes in \cite{dt}
since the induced charge background
field VEVs play the exact same role as the usual Higgs fields.

In the above, we show that to generate $M_{\rm GUT} = 10^{16}$~GeV,
we need the induced charged background to be
$\langle A \rangle = 10^{-2} M_{\rm S}$.
But it is usually not natural to generate the small shift
in the antisymmetric field strength $H$.
In fact, in the case that these
induced charged background VEVs satisfy $\langle F_{ij} \rangle = 0$,
they are naturally Wilson lines.
Since one has $\pi_1(K) = Z_N$ (where $K$ is the compact spacetime
manifold), we recover eq.~\ref{quant}.

In the following, we will propose a mechanism to make possible
small values of $\langle H \rangle$.
We assume that the compactification scale is at the gaugino
condensation scale rather than near the string scale, as is usually
assumed.
In this case,
\begin{equation}
\frac{1}{2\pi}\int_{\tilde{W}}H \sim R^3 H,
\end{equation}
where $R$ is the compactification scale, so $H \sim 1/R^3$, which is
small for large $R$.

We demonstrate this idea in the context of orbifold string models.
In the orbifold case, the quantization condition is better expressed in
that of the Wilson lines,
\begin{equation}
\oint A^i dx_i = \frac{2\pi}{n}.
\end{equation}
For orbifolds with the moduli background fields
\begin{equation}
G^{ij} = R^2 \delta^{ij},
\qquad G_{ij} = R^{-2} \delta_{ij},
\end{equation}
one has
\begin{equation}
\oint A^i dx_i = \oint A^i G_{ij} dx^i = \frac{2\pi A}{R} = \frac{2\pi}{n},
\qquad A = \frac{R}{n}.
\end{equation}
So in this scheme to have the small value for $A$, the
compactification radius is forced  to be large compared to the string scale.
Through the charged background, the compactification radius is related
to the gaugino condensation scale.
One can think of this as gaugino condensation inducing the compactification
of some dimensions and the charged background VEV's at the gaugino
condensation scale.

The fact that the compactification radius is on the same
order as the gaugino condensation scale makes the dynamics of
inducing the charged background VEV by the gaugino condensation
appear natural, but it may appear to invalidate our four-dimensional
description of gaugino condensation.
However, it has been proved perturbatively to all orders in string
perturbation theory \cite{provt} that string models are invariant under
$T$-duality \cite{dual}, {\em i.e.} $T \mapsto 1/T$ or $R \mapsto 1/R$.
In our case, $T$-duality implies our string models with large radius
can equally well be described by the theory with the small
compactification radius.
In the small-radius description, the gaugino condensation in the
hidden sector can be well approximated by the four-dimensional field
theory in which we know the gaugino condensation happens.
With the application of T-duality, we can see that the same dynamics
happens in our model.
This is a nice demonstration that $T$-duality is indeed a powerful tool!

We have argued above that there is a real hope for the existence of
affine level one grand unified string models with intermediate
scale gauge symmetry breaking.
But there are also some potential problems with this type of model
which we now discuss.

First of all, if one appeals to large moduli, then the string threshold
correction \cite{kap} to the gauge coupling constant could be
large and spoil the possible predictions of both  $M_{\rm GUT}$ and the
weak mixing angle in this scheme.
To avoid this problem, we can restrict attention to string models without
string threshold corrections. This kind of string models do exist.
It has been show that for string models with no sectors that preserve
$N = 2$ spacetime supersymmetry (for example, $Z_3$ and $Z_7$ orbifold
models) the gauge coupling constant does not receive moduli-dependent
string threshold correction \cite{dixon2,toms}.

Secondly, in this model, the gravitino gets large mass because of the
large gaugino condensation scale.
To generate the large mass hierarchy between $M_{\rm GUT}$ and $M_{\rm SUSY}$,
one should not break global supersymmetry in the observable sector.
This can be achieved if the string models have the no-scale structure which
have been shown \cite{wdr,rara} to occur in many string models.
The no-scale structure was originally proposed as a natural structure for
suppressing the cosmological constant at the tree level for the supergravity
theory \cite{noscale}.
Later, it was shown that in the no-scale models, when local supersymmetry
is broken, global supersymmetry is still preserved at tree level \cite{no}.
In the string models, the gauginos in the observable sector may still get
massive, because the gauge coupling constant is described by a function
of the dilaton and moduli fields.
However, one can easily show that for the string models with
moduli independent string threshold corrections, the observable sector
gaugino masses remain zero at tree level after local supersymmetry is broken.
Therefore, supersymmetry can be preserved in the observable sector even
though it is broken in the hidden sector.
The supersymmetry breaking may or may not be transmitted to the
observable sector through one-loop corrections \cite{gmass}.
We will defer analysis in this important question to future work.

Another interesting observation is that the dilaton runaway problem
may be solved in this scheme.
First, because of the induced small c-number, there
is no dilaton runaway problem at the tree level in these models;
the allowed small c-number also cures the quantization
 problem in the supersymmetry breaking scheme proposed in
ref.~\cite{wit2}.
Second, the usual arguments which lead to the conclusion that the
dilaton VEV is small do not hold in these models.
Because of the no-scale structure, the dilaton VEV should be determined
by the  radiative potential energy which typically yields
\begin{equation}
\label{typ}
\frac{e^{-(S + \bar{S}) / 2b}}{T+\bar{T}} \sim 1.
\end{equation}
(The fact that $S$ and $T$ appear in this combination
can be understood as a result of modular invariance.)
In the case, $T \sim 1$, dilaton VEV is $S \rightarrow 0$ which
corresponds to strongly-coupled string theory.
In the models proposed here, $T$ is determined by
\begin{equation}
ST \sim R^2 \sim e^{-(S+\bar{S}) / 2b}.
\end{equation}
It is not hard to see that this is consistent with
$\langle S \rangle \simeq 2$.
We will discuss the determination of the dilaton and moduli  VEVs
 in more detail in a future work.

In conclusion, we propose that gaugino condensation in the hidden sector
may dynamically induce intermediate scale gauge symmetry breaking and
specify the compactification radius.
This dynamical symmetry breaking scheme makes possible level one grand
unified string models.
It is interesting that $T$-duality plays an important role here although the
dynamics spontaneously breaks the symmetry.
We also show that the dilaton runaway problem can also be avoided in these
models.

\vskip 28pt
\noindent{\bf Acknowledgements}
\vskip 12pt

I would like to thank D. Arnowitt, M.K. Gaillard, H. Georgi,
S. Kelley, S. Kachru, J. Liu, and E. Witten
for helpful discussions.
\vskip 28pt

\vfill\eject

\end{document}